\documentclass[twocolumn,final,epj]{svjour}
\usepackage[utf8]{inputenc}
\usepackage[T1]{fontenc}
\usepackage{graphicx}
\usepackage{amssymb}
\usepackage{amsmath}
\usepackage{hyperref}
\usepackage{color}
\definecolor{blue-violet}{rgb}{0.54, 0.17, 0.89}
\definecolor{ao(english)}{rgb}{0.0, 0.5, 0.0}

\begin{document}

\title{Quantum chaos of dark matter in the Solar System}

\author{
D.L. Shepelyansky\inst{1}}

\institute{
Laboratoire de Physique Th\'eorique du CNRS, IRSAMC, 
Universit\'e de Toulouse, CNRS, UPS, 31062 Toulouse, France
}

\date{Dated: 17 Dec 2017; updated: January 22, 2017}

\abstract{We perform time-dependent analysis of quantum dynamics of dark matter 
particles in the Solar System. It is shown that this problem
has similarities with a microwave ionization of Rydberg atoms
studied previously experimentally and analytically.
On this basis it is shown that the quantum effects for
chaotic dark matter dynamics become significant
for dark matter mass ratio to electron mass
being smaller than $2 \times 10^{-15}$. Below this border a multigraviton 
(analogous to multiphoton) diffusion over Rydberg states of dark matter atom
becomes exponentially localized in analogy with the Anderson
localization in disordered solids. The life time
of dark matter in the Solar System is determined
in dependence on mass ratio in the 
localized phase and a few graviton ionization regime
(analogous to a few photon ionization in atomic physics).
The quantum effects for dark matter 
captured by other binary systems are also discussed.  
}


\maketitle

\section{Introduction}

The properties of dark matter 
are now actively discussed by the astronomy community
(see e.g. \cite{bertone2005}). Recently, a necessity of correct
description of galactic structures, in particularly 
singular density cusp problem,  attracted 
a growing interest to  the ultralight dark matter 
particles (DMP) of bosons with a mass 
$m_d \sim 10^{-22} eV$ (see e.g.  \cite{marsh,calabrese2016,hu2016,lee2017}
and Refs. therein). 
However, the mass $m_d$ of light DMP
is unknown and possibilities of its detection
are under active discussions \cite{marsh,sibiryakov}.
At small values of $m_d$ (or its ratios to electron mass $m_e$)
the quantum effects start to be dominant \cite{hu2016,lee2017,gruzinov}.
Till present the quantum effects have been studied 
in the frame of static solutions of Schr\"odiner and Poisson equations.

In this work we perform a time-dependent analysis of quantum effects
for light DMP which dynamics takes place in binary 
rotating systems.
The properties of quantum dynamics of such DMP
in the Solar System (SS), its 
atomic Rydberg structure (similar to hydrogen atom) 
and multiphoton ionization (escape from SS)
are analyzed here for SS and more generic binary systems.
For the SS we consider the model
of Sun with Jupiter which creates a time periodic perturbation
leading to dynamical chaos and  diffusion of DMP energy
with eventual escape (or ionization) from the SS.
In the quantum case the periodic gravitational force, 
created by rotation of Jupiter,
leads to absorption or emission of photons changing the
 DMP energy by integers of $\hbar \omega_J$ 
where the frequency $\omega_J=2\pi/T_J$
is determined by the period of Jupiter rotation.
We call these  energy quanta as photons even if there 
are no electromagnetic forces in this system
(a reader can also look on these quanta as gravitons).

Here we should note that below we speak about photon transitions
to stress analogy of quantum DPM effects with a process of
microwave ionization of Rydberg atoms.
However, formally for quantum DMP in the Solar System
of Sun and Jupiter there are only gravitational forces
acting on DMP. Thus formally one should say that the time-dependent rotation
of Jupiter generates gravition absorption or emission
of gravitons (analogous to photons in atomic physics)
with energies being integers of $\hbar \omega_J$
in DMP energy variation.
In further we keep the notations of
atomic physics using the term {\it photon}
as an equivalent of {\it graviton}.

Below we show that  for a light DMP mass with $m_d/m_e < 2 \times 10^{-15}$
the quantum effects start to play a dominant role and that they
lead to a dynamical localization of diffusive chaotic motion of DMP
in binary system being similar to 
the Anderson localization in disordered solids
(see \cite{anderson1958,akkermans2007,evers2008}).  
Due to localization a DMP escape 
from SS is strongly suppressed and DMP life time in SS 
is increased enormously.
A similar dynamical localization of chaotic diffusion
of multiphoton transitions has been predicted
for microwave ionization of excited hydrogen and Rydberg atoms 
and observed in experiments 
(see \cite{hydrogen1,hydrogen2,kochexp,koch,rydberg} and Refs. therein).
We show that the DMP ionization from SS induced by Jupiter
has many similarities with physics of 
multiphoton ionization of atoms in strong laser fields \cite{delone}
and properties of Rydberg atoms in external fields \cite{gallagher}.
Since the classical DMP dynamics in SS 
is mainly chaotic the quantum evolution of DMP
has many properties of quantum chaos \cite{haake}. 
We show that one of the consequences of classical and quantum chaos 
in binary systems is an absence of 
singular density cusp in center of a binary.

\section{Kepler map description of classical DMP dynamics}

We consider the restricted three-body problem \cite{3body} with 
a DMP of light mass $m_d$, Sun of mass M and a planet (Jupiter) of mass $m_p$
moving around Sun over a circular orbit of radius $r_p$ with velocity $v_p$ and
frequency $\omega_p = v_p/r_p$. For the Jupiter case we have
$v_J=v_p = 13.1 km/s$, $r_J=r_p = 7.78 \times 10^{8} km = 5.204 ASTRU$ 
(ASTRU is for astronomy units), 
orbital period $T_p =2\pi/\omega_p = 11.8 yrs$ and $m_p/M=1/1047$ 
\cite{wikijupiter}.
The studies of DMP dynamics in a binary system with $m_p \ll M$ showed that 
the dynamics of comets or DMP 
is well described by the generalized Kepler map
 \cite{petrosky,halley,duncan,tremaine1999,shevchenko,lages2013,rollin2015a,rollin2015b}:
\begin{eqnarray}
E_{n+1} &=& E_n+F(\phi_n) \; , \\ 
\nonumber
\phi_{n+1} &=& \phi_n + 2\pi  |2E_{n+1}/(m_d {v_p}^2)|^{-3/2} \; ,
\label{eq1}
\end{eqnarray}
where $E_n$ is DMP energy, $\phi_n$ is Jupiter phase
taken at n-th passage of DNP through
perihelion on a distance $q$ from Sun.  
This symplectic map description
is well justified for $q > r_p$ where the kick function 
$F(\phi) = f_0 (m_p/M) m_d {v_p}^2 \sin \phi $ and 
$f_0 \approx 2 (r_p/q)^{1/4} \exp(-0.94 (q/r_p)^{3/2})$.
For $q \sim r_p$, like for the comet Halley case, the function $F(\phi)$
contains also higher harmonics 
with a maximal kick amplitude $f_0 \approx 2.5$ for the comet Halley
\cite{halley,rollin2015a}. The map is valid when the orbital DMP period
is larger than the planet period. This map generates a chaotic DMP dynamics
similar to those of the Chirikov standard map  \cite{chirikov1979,lichtenberg}
for energy being below the chaos border 
$|E| < E_{ch} =  w_{ch} m_d {v_p}^2/2$ with  
$w_{ch} \approx 2.5 (2 f_0 m_p/M)^{2/5}$. Examples of Poincar\'e sections
for the generalized Kepler map are given in \cite{halley,rollin2015a,rollin2015b}.
For the case of Jupiter with $\sin-$kick function
we have $w_{ch} \approx 0.3$ while for the comet Halley case
with a few harmonics one finds $w_{ch} \approx 0.45$.
In the chaotic phase the energy is growing in a diffusive way
with number of DMP orbital periods $t_{orb}$: $(\Delta E)^2 \approx D t_{orb}$. 
The diffusion coefficient $D$
is approximately given by the random phase approximation for phase $\phi$:
\begin{equation}
\label{eq2}
D \approx <F^2(\phi)> \approx {f_0}^2 (m_p/M)^2 {m_d}^2  {v_p}^4/2 \; .
\end{equation}

For a DMP orbit with initial energy about $-m_d {v_p}^2/2$  the ionization energy
is $E_I = m_d v_p^2/2$ and a diffusive ionization time (escape from SS) is approximately
$t_D \approx 2\pi r_p{E_I}^2/(v_pD) $ that gives for SS with Jupiter
$t_D \approx 3 \times 10^6 yrs$. More detailed numerical simulations with many DMP
trajectories, including the case of comet Halley, give  
a typical time scale
$t_I \approx t_H \approx 10^7 yrs \sim t_D$  \cite{halley,lages2013}.  
For an external galactic  DMP flow
scattering on  SS, DPM are captured and accumulated during the time scale $t_H$.
After this time scale the DMP distribution in SS reaches a steady-state when 
the capture process is compensated by escape on a time scale $t_I$.
The capture process and its cross-section are discussed in detail in
\cite{khriplovich,peter,lages2013}.

It is also useful to note that a rotating planet corresponds to
a rotating dipole in the Coulomb problem
which can be transformed to a circular polarized
monochromatic field appearing in 
the problem of microwave ionization of Rydberg atoms 
and autoionization of molecular Rydberg states \cite{benvenuto}.

\section{``Hydrogen'' atom of dark matter}

Since the gravitational interaction is 
similar to the Coulomb interaction we directly obtain from \cite{landau} 
the levels of dark matter 
atom with the Bohr radius 
$a_{Bd} =  \hbar^2/(\kappa {m_d}^2 M) = 1.01 \times 10^{-26} cm (m_e/m_d)^2$ for SS and energy levels
$E_{dn} = - E_{Bd}/(2 n^2)$  where $\kappa$ is the gravitational constant.
Here we have the dark matter atomic energy  $E_{Bd}=  \kappa m_d M/a_{Bd}
= 7.47 \times 10^{36} (m_d/m_e)^3 ev$ and
related atomic frequency 
$\omega_{Bd} = E_{Bd}/\hbar = \kappa^2 {m_d}^3 M^2/\hbar^3$.
For the SS we have the Bohr radius 
$a_{Bd}  = 1.30 \times 10^{-40} (m_e/m_d)^2 r_J$
so that $a_{Bd}=r_J$ at $m_d/m_e =  1.14 \times 10^{-20}$.
Thus we have  
$E_{Bd} = 1.10 \times 10^{-23} ev = \hbar \omega_J$
when  $a_{Bd}=r_J$
and  one photon energy of Jupiter frequency $\omega_J=\omega_p = v_p/r_p$
is $\hbar \omega_J$. Hence  one needs 
$N_J = E_{Bd}/(2 \hbar \omega_J) 
= 3.38 \times 10^{59} (m_d/m_e)^3 = 0.5$
photons to ionize the ground state of DMP atom at this mass ratio.

\section{Quantum Kepler map and Anderson localization}

In analogy with the microwave ionization
of excited hydrogen and Rydberg atoms \cite{hydrogen2,rydberg}
the quantum dynamics of DPM is described by the quantum Kepler map
obtained from (\ref{eq1}) by replacing the classical variables $(E,\phi)$
by operators $\hat{E}=\hbar \omega_p \hat{N}$, $\hat{\phi}$ with a commutator
$[\hat{N},\hat{\phi}]=-i$. Here $N$ has the meaning of a number of photons
absorbed or emitted due to interaction with periodic perturbation of planet.
The quantum Kepler map has the form \cite{hydrogen2,rydberg}:
\begin{equation}
\label{eq3}
 \hat{{\bar N}} = \hat{N} + k \sin \hat{\phi} \; , 
\hat{{\bar \phi}} = \hat{\phi} + 2\pi \omega  (- 2 \omega \hat{{\bar N}})^{-3/2} \; ,
\end{equation}
where bars mark new values of operator variables after one orbital period of DMP
and a kick amplitude $k= f_0 m_p m_d {v_p}^2/(\hbar \omega_p M) 
=2 f_0 (m_p/M) N_I = 2.10 \times 10^{17} (m_d/m_e)$ gives
the maximal number of absorbed/emitted photons after one kick
(numbers are given for Jupiter case). 
Here we express the planet frequency $\omega_p$ in atomic units of dark matter atom
with $\omega = \omega_p/\omega_{Bd}=1.48 \times 10^{-60} (m_e/m_d)^3 $. 
The corresponding wavefunction evolution in the basis of photons $N$ is described by
the map which is similar to the quantum Chirikov standard map \cite{hydrogen2,rydberg}:
\begin{equation}
\label{eq4}
\bar{\psi_{N_\phi}} = \exp(-k\cos \phi) \exp(-i H_0(N_\phi)) \psi_{N_\phi}
\end{equation}
with $H_0(N_\phi) = 2\pi (-2 \omega (N_0 + N_\phi))^{-1/2}$ and
$N=N_0+N_\phi$,
where $N_0$ is the number of photons of DMP initial state.
For the initial DMP state with energy $E_d = - m_d v_p^2/2$
we have the number of photons
required for DMP ionization being 
\begin{equation}
\label{eq5}
N_I = m_d v_p^2/(2 \hbar \omega_p) =  4.39 \times 10^{19} m_d/m_e
\end{equation}
with $N_0= -N_I$  and the right equality given for the Jupiter case at $f_0=2.5$.
In this expression for $N_I$ we use $w_{ch} \approx 1$ 
and assume that $a_{Bd} < r_J$. For $a_{Bd} > r_J$ the minimal energy
of DMP is given by the ground state $E = -E_{Bd}/2$.

\begin{figure}[h]
\begin{center}
\includegraphics[width=0.46\textwidth]{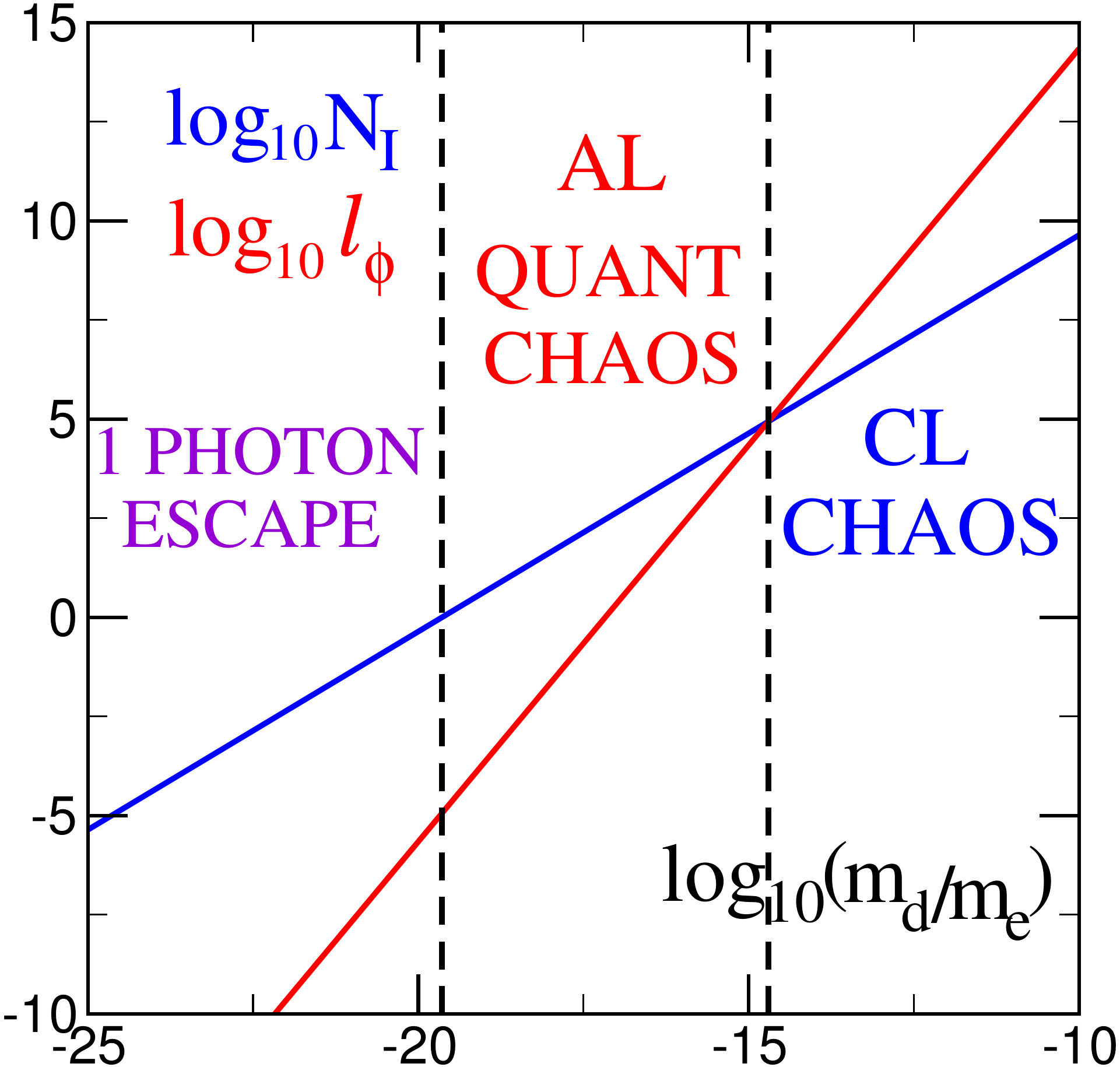}
\caption{Dependence of number of photons $N_I$ of Jupiter
frequency required for ionization (escape) of DMP
on its mass $m_d$ rescaled by electron mass $m_e$ (blue line
given by the analytic expression (\ref{eq5})),
dependence of Anderson localization length of quantum chaos
$\ell_\phi$ on $m_d/m_e$ (red line
given by the analytic expression (\ref{eq6}));
vertical dashed lines mark regimes of different 
ionization mechanisms of DMP:
one photon escape (left), Anderson localization (middle),
classical chaos with diffusive ionization (right);
here initial DMP energy is $E_d=-m_d {v_J}^2/2$.}
\label{fig1}
\end{center}
\end{figure}

The quantum interference effects lead to 
exponential localization of chaotic diffusion 
being similar to the Anderson localization in disordered solids
\cite{akkermans2007,evers2008}. 
In analogy with the microwave ionization of hydrogen atoms,
the localization length expressed in the
number of photons is \cite{hydrogen2,rydberg}:
\begin{eqnarray}
\nonumber
 \ell_\phi &\approx&   D/(\hbar \omega_p)^2 \approx k^2/2 
     = 2{f_0}^2(m_p/M)^2 {N_I}^2 \\ 
\nonumber
     &\approx&  \; {f_0}^2 {m_p}^2 {m_d}^2 {v_p}^4/2(\hbar \omega_p M)^2 \\
     &\approx& 2.20 \times10^{34} (m_d/m_e)^2
\label{eq6}
\end{eqnarray}
where the last equality is given for the Jupiter case at $f_0=2.5$.

The wavefuction $\psi_{N_\phi}$ is exponentially localized
giving a steady-state probability distribution over photonic states:
\begin{eqnarray}
<|\psi_{N_\phi}|^2> &=& W(N_\phi) 
\label{eq7} \\
 \approx  (1+2|N_\phi|/\ell_\phi) 
&\exp&(-2|N_\phi|/\ell_\phi)/2\ell_\phi .
\nonumber 
\end{eqnarray}
The above expression for $\ell_\phi$ is valid for $\ell_\phi > 1$ while 
for $\ell_\phi < 1$ we enter in the regime of perturbative localization.
The steady-state localized distribution (\ref{eq7})
is settled on a quantum time scale 
$t_q \approx T_p \ell_{\phi}$ \cite{hydrogen1,hydrogen2,rydberg}.

The localization takes place for the photonic range $|N_\phi| < N_I$
and it is well visible for $\ell_\phi < N_I$. For $\ell_\phi > N_I$
a delocalization takes place and DMP escape is well described by 
the classical chaotic dynamics
and diffusion. For Jupiter case and DMP at initial energy $E_d = -  m_d v_p^2/2$,
with the corresponding $N_I$ (we assume here the chaos border
$w_{ch} \approx 1$, for $w_{ch} < 1$ we should multiply $N_I$ by $w_{ch}$). Thus we find that 
the delocalization takes place at 
\begin{equation}
\label{eq8}
m_d/m_e > \hbar \omega_p (M/m_p)^2/({f_0}^2 m_e {v_p}^2) = 2 \times 10^{-15} \; ,
\end{equation}
with the last equality given for the Jupiter case.
For smaller ratios $m_d/m_e < 2.0 \times 10^{-15}$ we
have Anderson like localization
of DMP probability on the photonic lattice.
At the delocalization border with $\ell_\phi = N_I$ we have
$N_I=8.78 \times 10^4$ at the above value of $m_d/m_e$
so that the DMP ionization goes via highly multiphoton process.

An example of probability distribution for a localized state at $\ell_\phi = 1.39$ 
and $m_d/m_e = 7.95 \times 10^{-18}$ corresponds  to Fig.3 in \cite{rydberg}.
The number of photons $N_I$ required for ionization of an initial DMP state
with energy $E_d$ also depends on the mass ratio $m_d/m_e$
so that one photon ionization takes place for $N_I < 1$
corresponding to $m_d/m_e < 2.28 \times 10^{-20}$.

The different regimes of quantum DMP dynamics are shown in Fig.~\ref{fig1}.
The classical description is valid for $\ell_\phi > N_I$
corresponding to $m_d / m_e > 2 \times 10^{-15}$, 
the Anderson photonic localization
takes place in the range $2.28 \times 10^{-20} < m_d / m_e < 2 \times 10^{-15}$
and one photon ionization appears for $m_d / m_e < 2.28 \times 10^{-20}$.

Even if the quantum Kepler map gives an approximate 
description of quantum excitation, it was shown that
it provides a good description \cite{qkeplermap} of microwave ionization
of real three-dimensional excited hydrogen atoms
in delocalized and localized regimes \cite{koch,kochexp}.
This result justifies the given description of quantum DMP dynamics
in SS.

\section{Ionization times}

The typical time scales of ionization (escape) for these three regimes
can be estimated as follows. In the delocalized phase  $\ell_\phi > N_I$ 
the ionization time is determined by 
a diffusive process with $t_I \approx t_H \approx 10^7 yrs$
as obtained from extensive numerical simulations of comet Halley \cite{halley}
and classical chaotic dynamics of DMP \cite{lages2013}. This regime $t_I $ 
is independent of $m_d$.

In the localized phase $\ell_\phi < N_I$ the ionization takes place only
from the exponentially small tail of the steady-state probability
distribution (\ref{eq7}) with the escape rate 
$\Gamma \sim W \sim (N_I/\ell_\phi) \exp(-2|N_I|/\ell_\psi)$
so that we obtain  the estimate for ionization  time $t_I \sim 1/\Gamma$:
\begin{equation}
\label{eq9}
t_I \approx t_H  \exp(2|N_I|/\ell_\phi-2) /(2|N_I|/\ell_\phi-1) \; .
\end{equation}
The above expression assumes that $\ell_\phi > 1$ and $N_I \geq \ell_\phi$
giving $t_I = t_H $ at delocalization border $N_I=\ell_\phi$.

\begin{figure}[h]
\begin{center}
\includegraphics[width=0.46\textwidth]{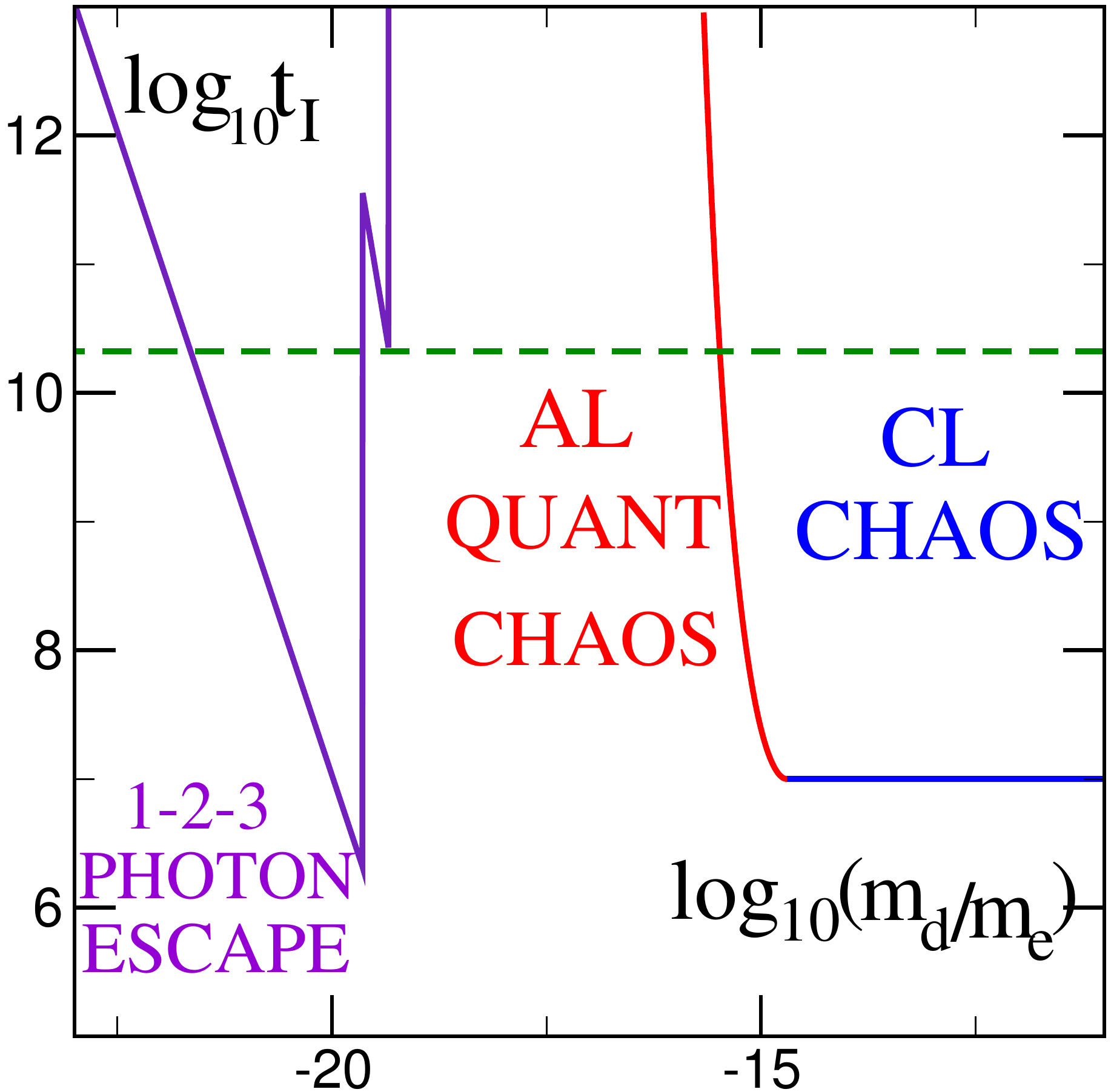}
\caption{Dependence of DMP escape (ionization) time $t_I$
from the SS (expressed in years)
as a function of mass ration $m_d/m_e$
for initial DMP energy $E_d= - m_d {v_J}^2/2$.
The blue horizontal line
shows regime of classical escape due to chaos
($t_I \approx 10^7$ years independent of $m_d/m_e$),
red curve shows $t_I$ of the analytic expression
(\ref{eq9}) in the regime of 
Anderson localization of quantum photonic
transitions, violet line shows 
the analytic expression  (\ref{eq10}) for $t_I$
in the regime of 1,2 and 3 photon escape;
horizontal dashed line marks the life time of Universe $t_U$.
}
\label{fig2}
\end{center}
\end{figure}

In the case when $N_I < 1$ an absorption of one photon with energy 
$\hbar \omega_p$
is sufficient to give to DMP positive energy leading 
to its escape on infinity or ionization.
In this regime the one-photon ionization rate is 
$\Gamma \approx (\omega_p/2\pi) (J_1(k))^2$ where $J_1(k)$ is Bessel function.
This simple estimate, following from the quantum Kepler map (\ref{eq4}), 
is in a good agreement with the exact computation
of one-photon ionization rate as it is demonstrated in \cite{hydrogen1,hydrogen2}.
Hence, the one-photon ionization time is 
\begin{equation}
\label{eq10}
t_I \approx 1/\Gamma \approx T_p (2/k)^2 \approx 1.07 \times 10^{-33} (m_e/m_d)^2 yrs \; .
\end{equation}
Thus at one-photon border $N_I=1$ and $m_d/m_e= 2.28 \times 10^{-20}$
we have $t_I = 2.05 \times 10^6 yrs$. 
For $2.28 \times 10^{-20} < m_d/m_e <4.56 \times 10^{-20}$
two photons are required for DMP ionization with
$\Gamma =   (\omega_p/2\pi) (J_2(k))^2 $ and  slightly above
one photon ionization border we obtain
$t_I \approx T_p  (2/k)^4 \approx 3.6 \times 10^{12} yrs$  
being much larger than the age of universe
$t_U \approx 1.38 \times 10^{10} yrs$ (at the 3-photon border we have
$t_I \approx  4 (2/k)^6 T_p \approx 4 \times 10^{15} yrs$).

The global dependence of escape time $t_I$ on DMP mass $m_d$
is shown in Fig.~\ref{fig2}. The life time is larger than
the life time of Universe for the mass ratio
$ 2.2 \times 10^{-20} < m_d/m_e < 3.4 \times 10^{16}$
where the left inequality is at the transition from 2-photon
to 1-photon ionization. For $m_d/m_e < 2.8 \times 10^{-22}$
the one-photon ionization becomes very slow and we also have
$t_I > t_U$. In this range we have the atomic size of
DMP atom $a_{Bd} \gg r_J$ and the above ionization time
is given for ionization from the ground state.
The time $t_I$ for one-photon process 
becomes so large because 
in the quantum Kepler map (\ref{eq4}) the kick amplitude 
$k \propto m_d$ becomes very small. 

It would be interesting to verify the above estimates for ionization
rates $\Gamma$ and life times $t_I$ by the numerical simulation
methods developed for computation of these quantities 
in microwave ionization of Rydberg atoms \cite{delande}.

\section{DMP capture}

For the classical DMP of Galactic wind flying through SS
the capture cross-section is 
$\sigma \approx 8\pi {r_p}^2 (v_p/v)^2$ 
being diverging at low positive DMP energies $E=m_d v^2/2 >0$ in a continuum
\cite{khriplovich,lages2013}. The captured DMP diffuse in the chaotic 
region up to the chaos border $w_{ch} = 2|E|/(m_d {v_p}^2)$ as discussed above
(see also \cite{lages2013}).
Due to Anderson photonic localization (\ref{eq6})
the diffusion is localized and,
comparing to the classical border $w_{ch}$, DMP can reach only significantly smaller
quantum border values:
\begin{equation}
\label{eq11}
w_q  \approx 2 \hbar \omega_p \ell_\phi/(m_d {v_p}^2) 
\approx  5. \times 10^{14} (m_d/m_e) \; .
\end{equation}
This dependence is valid in the range 
$2.28 \times 10^{-20} < m_d/m_e < 2 \times 10^{-15}$.
For $m_d/m_e$ becoming smaller than the left inequality we have 
$\ell_{\phi} < 1$ and only one photon energy is absorbed with
$w_q = 1.14 \times 10^{-5}$; above the right border we obtain the classical
chaos border with $w_q \approx w_{ch} \sim 1$ independent of DMP mass.

According to analysis given in \cite{lages2013}
the classical capture process of DMP continues during time 
$t_H \sim t_D$ after which DMP start to escape from SS.
Assuming that the Galactic DMP velocity distribution
has a usual Maxwell form $f(v) dv = \sqrt{54/\pi} v^2/u^3 \exp(-3v^2/u^2) dv$
with $u \approx 220 km/s$ we can estimate the captured DMP mass $\;\;\;$ as \cite{lages2013} :
$M_{cap} \sim 100 (v_p/u)^3 (m_p/M) \rho_g {r_p}^2 v_p t_H$,
where $\rho_g \approx 4 \times 10^{-25} g/cm^3$ is the Galactic mass density  
of dark matter. In the quantum case with photonic localization
we should use $t_q < t_H$ since the accumulation continues only during time
on which a steady-state distribution is reached while after it
the escape of DMP from SS starts to compensate ingoing DMP flow.
Thus $M_{cap}$ is significantly reduced by the factor $t_q/t_H$.

In the above estimate we have $M_{cap} \propto E_H \propto m_d {v_p}^2 w_H /2$ 
where $w_H \sim 2f_0(m_p/M)$ and $E_H$ has a meaning of 
DMP energy which can be captured by a kick from Jupiter. 
In the one-photon regime with $k<1$ at $m_d/m_e < 4.76 \times 10^{-18}$
only DMP energies with $E=m_d v^2/2 < \hbar \omega_J$ 
can be captured that provides an additional reduction of $M_{cap}$.

Finally, the Kepler map approach allows to
perform extensive simulations of DMP capture process 
for SS and other binaries up to time scales
of SS life time \cite{lages2013,rollin2015b}
with a steady-state DMP distribution
reached at such times. The obtained 
DMP steady-state distribution has maximal 
volume density on a distance comparable
with a size of binary without 
cusp singularity at $r \ll r_p$. 
The physical reason is rather clear:
DMP diffuse only up to the chaos border $w_{ch} \sim 1$
corresponding to hallo distances from the binary center 
$r_h \sim r_p/w_{ch} \sim r_p$. In the regime of quantum localization
we should replace $w_{ch}$ by $w_q < w_{ch}$
so that $r_h \sim r_p/w_q \gg r_p$ becomes even larger.
Thus in presence of time-dependent effects in binaries 
there is no cusp singularity at the binary center,
both in classical and quantum cases.

\section{Other binary example}

Above we considered the binary case of Sun and Jupiter.
However, the obtained results can be directly extended to 
binary systems with other parameters.
As an example we consider  Sagittarius $A^*$ which is thought 
to be the location of a supermassive black hole (SBH)
in our Galaxy \cite{wikigalaxybh} with a mass $M_{bh} \approx 4 \times 10^6 M_S$.
The nearby star $S2$ moving around SBH has the following characteristics \cite{wikigalaxys2}:
the orbital period $T_{s2} \approx 15yrs \approx 1.3 T_p$ is comparable with the period of
Jupiter $T_p$, the semi-major axis is $r_{s2}  = 980 ASTRU = 188 r_J$
and mass $m_{s2} = 15 M_S$. 
Hence $\omega_{s2}/\omega_J \approx  0.7$ and the velocity of $S2$ is
$v_{s2}/v_J \approx  144 $ or $v_{s2} \approx 1900 km/s$.
Other stars around SBH are assumed to give a small perturbation,
at present there masses are not know exactly  \cite{wikigalaxybh}.
With these parameters of binary composed by SBH  Sagittarius $A^*$ and the star $S2$ 
(assuming regime of chaotic dynamics)
we obtain
from (\ref{eq8}) that the quantum effects of Anderson photonic localization start
to play a role for the mass ratio $m_d/m_e < 1.7 \times 10^{-14}$,
being a bit higher than for the case of Sun and Jupiter.

\section{Discussion}

We performed analysis of
time-dependent effects for DMP quantum dynamics in 
binary systems. On the basis of results obtained for multiphoton
ionization of Rydberg atoms we show the emergence of
Anderson photonic localization for DMP with masses
$m_d/m_e < 2 \times 10^{-15}$. The life times
of DMP in SS are determined in the localized regime and
a few photon quantum ionization regime.
The obtained results determine the life time
of DMP in a binary systems as a function of
DMP mass and masses of binary. 
The probability distribution for DMP over energy 
is obtained for the Anderson localization regime
and the capture process characteristics are determined.
These features can be accessible for detectors of dark matter. 


This work is supported in 
part by the Programme Investissements
d'Avenir ANR-11-IDEX-0002-02, 
reference ANR-10-LABX-0037-NEXT (project THETRACOM).


\end{document}